\documentclass[10pt,a4paper]{article}

\usepackage[T1,T2A]{fontenc}
\usepackage[utf8]{inputenc}
\usepackage[russian, english]{babel}

\usepackage[affil -it]{authblk}

\usepackage{graphicx}
\usepackage{amsmath,amsfonts,amssymb,textcomp}
\usepackage{color}
\usepackage{xcolor}

\usepackage{amsmath}

\usepackage[sorting=none, style=phys, backend=biber, uniquename=init, giveninits=true]{biblatex}
\bibliography{references}

\usepackage[colorlinks]{hyperref}
\usepackage[colorinlistoftodos, textsize=tiny, textwidth=110,]{todonotes}
\setuptodonotes{fancyline}

\usepackage{array,multirow,tabularx, makecell}
\usepackage{adjustbox}

\usepackage[font=scriptsize]{caption}

\usepackage{sectsty}
\sectionfont{\centering}

\usepackage{indentfirst}

\title{\textbf{Cross Sections of Solar Neutrino Capture by $^{127}$I Nuclei and Gamow–Teller Resonances}}
\author[1,*]{ Yu.\,S.\, Lutostansky}
\author[1,2,3]{A.\,N.\,Fazliakhmetov}
\author[1,2,**]{G.\,A.\, Koroteev}
\author[1]{N.\,V.\, Klochkova}
\author[1]{A.\, P.\, Osipenko}
\author[1]{ V.\,N.\, Tikhonov}

\affil[1]{National Research Centre "Kurchatov Institute", Moscow, Russia}
\affil[2]{Moscow Institute of Physics and Technology, Dolgoprudny, Russia }
\affil[3]{Institute for Nuclear Research of Russian Academy of Sciences, Moscow, Russia}

\affil[ ]{ }
\affil[ ]{E-mail:}
\affil[*]{ lutostansky@yandex.ru}
\affil[**]{ koroteev@phystech.edu}

\begin{document}

\maketitle

\begin{abstract}
Solar neutrino capture cross-section $\sigma(E)$  by $^{127}$I nucleus has been studied with taking into account the influence of the resonance structure of the nuclear  strength function $S(E)$. 
Three types of isobaric resonances: giant Gamow-Teller, analog resonance and low-lying Gamow-Teller pigmy resonances has been investigated on the framework of  self-consistent theory  of finite Fermi systems.
The calculations  have been performed considering the resonance structure of the charge-exchange strength function $S(E)$. We analyze the effect of each resonance on the energy dependence of $\sigma(E)$.
It has been shown that all high-lying  resonances should be considered in the  $\sigma(E)$ estimations. 
Neutron emission process for high energy nuclear excitation leads to formation $^{126}$Xe isotope. 
We evaluate contribution from various sources of solar neutrinos to the  $\sigma(E)$ structure and to the $^{126}$Xe/$^{127}$Xe isotopes ratio formed by energetic neutrinos.
$^{126}$Xe/$^{127}$Xe isotope ratio could be an indicator of high-energy \emph{boron} neutrinos in the solar spectrum. 
We also discuss the uncertainties in the often used Fermi-functions calculations. 



\end{abstract}

\section{Introduction}
In neutrino physics and astrophysics, the process of interaction of neutrino with matter is of great importance.
In most cases, it is needed to calculate neutrino capture cross-section $\sigma(E)$ and take into account the structure of the charge-exchange strength function $S(E)$, which determines the magnitude of $\sigma(E)$ and its energy dependence.
The charge-exchange strength function $S(E)$ has a resonant nature, and its structure affects the neutrino capture cross-section $\sigma(E)$.
This is particularly important for the simulation of neutrino detectors based on the $\nu$-capture reaction
\begin{equation} \label{eq:nu_A(NZ)capture}
    \nu_{e} + A(N,Z) \to e^{-} + A(N-1,Z+1)
\end{equation}
The first proposed isotope for the neutrino detection was $^{37}$Cl.
The chlorine–argon radiochemical method was proposed by Pontecorvo in 1946 \cite{PontecorvoReport} and later implemented by Davis \cite{Davis_PhysRevLett.12.303} in the United States.
Implementation of the gallium–germanium method for measuring solar neutrinos was started after \cite{KuzminReport}, \cite{KUZMIN196527} papers.
A low-threshold detector for the $^{71}$Ga$(\nu_e, e^{-}) ^{71}$Ge reaction was established at an underground laboratory and was well protected from cosmic rays (for more details, see review \cite{Ryazhskaya_2018}).

Another approach of neutrino detection is iodine-xenon reaction based radiochemical method
\begin{equation} \label{eq:127I_capture}
    \nu_{e} + ^{127}\mathrm{I} \to e^{-} + ^{127}\mathrm{Xe}
\end{equation}
with the threshold of $Q= 662.3 \pm 0.20 $~keV  lower than the $^{37}$Cl one ($Q=813.87 \pm 0.20$~keV \cite{Wang_2017}), which increases the neutrino capture cross-section $\sigma(E)$.
Such detector can detect not only $^8$B solar neutrinos but also $^7$Be ones.
In 1988, Haxton \cite{Haxton_PhysRevLett.60.768} pointed out that the cross-section $\sigma(E)$ for the reaction (\ref{eq:127I_capture}) on iodine could be much larger than that on $^{37}$Cl and that the detection volume for iodine can be made much larger than the chlorine detector.
However, no calculations of the cross-section $\sigma(E)$ for the reaction  (\ref{eq:127I_capture}) were performed at that time, only estimations.
In the following year, calculations in \cite{Lutostansky_Shulgina_Report} were performed including the resonance structure of the charge-exchange strength function $S(E)$ of the  $^{127}$Xe daughter nucleus.
In 1991, those calculations were refined in \cite{Lutostansky_Shulgina_PhysRevLett.67.430} with allowance for special features of the normalization of the strength function $S(E)$, and the \emph{quenching}-effect was taken into account (for more details, see \cite{Lutostansky_Tikhonov2018}).
In the late 90s, there was a proposal to develop solar neutrino detector with  $^{127}$I as the target in the Homestake laboratory (USA) \cite{homestake1}, \cite{homestake2}, \cite{lande1}. 
Also in Los Alamos National laboratory neutrino capture cross-section with  $^{127}$I as the target was investigated on the accelerator beam   \cite{LosAlamos_cross_sec}.
The strength function $S(E)$ was measured in the $^{127}$I$(p,n) ^{127}$Xe reaction in 1999 \cite{Palarczyk_PhysRevC.59.500}, and these experimental data appeared to be in good agreement with our predictions made in \cite{Lutostansky_Shulgina_Report} \cite{Lutostansky_Shulgina_PhysRevLett.67.430}.
The calculations performed at that time by Engel, Pittel, and Vogel  \cite{Engel_Pittel_Vogel_PhysRevLett.67.426} \cite{Engel_Pittel_Vogel_PhysRevC.50.1702} are also noteworthy.
According to \cite{Palarczyk_PhysRevC.59.500}, the comparison with the experimental dependence of the strength function $S(E)$ demonstrates that the calculations in \cite{Lutostansky_Shulgina_PhysRevLett.67.430} have a higher prediction accuracy.

However, previous calculations did not include the possibility of the formation of the stable $^{126}$Xe isotope in the neutrino capture reaction (\ref{eq:127I_capture}) in the iodine detector.
High-lying charge-exchange resonances in the strength function $S(E)$ determine the formation of the stable $^{126}$Xe isotope upon the capture of high energy solar neutrinos by the $^{127}$I nucleus and the subsequent emission of a neutron from the formed $^{127}$Xe nucleus.

In the resonance structure of the charge-exchange strength function of  $^{127}$Xe nucleus can be identified three types of isobaric resonances (see Fig. \ref{fig:1}): giant Gamov-Teller (GTR), analog (AR), and low-lying Gamov-Teller pigmy resonances (PR) \cite{Lutostansky2017_JETP}.
Similar resonance structure can also be observed in other neutron-rich nuclei \cite{Lutostansky2018_EPJ}.
Recently it was shown \cite{Lutostansky_Tikhonov2018} \cite{Lutostansky2019_PhysicsOfAtomicNuclei} \cite{Lutostansky:2019iri} that in the calculation of the total neutrino capture cross-section $\sigma(E)$ should take into account all types of resonances. 
Ignoring even high-lying ones, such as GTR, leads to a deficit in the cross-section $\sigma(E)$, which can significantly affect the interpretation of experimental data.
In calculations of $\sigma(E)$ for capture solar neutrinos \cite{Lutostansky:2019iri} \cite{LutostanskyEtAl2020_PhysicsOfAtomicNuclei} were obtained similar results demonstrating that it is necessary to include all resonances in the charge-exchange strength function $S(E)$.

\section{Excited states structure of the $^{127}$Xe nucleus}
\begin{figure}[ht!]
\centering
\includegraphics[width=0.7\linewidth]{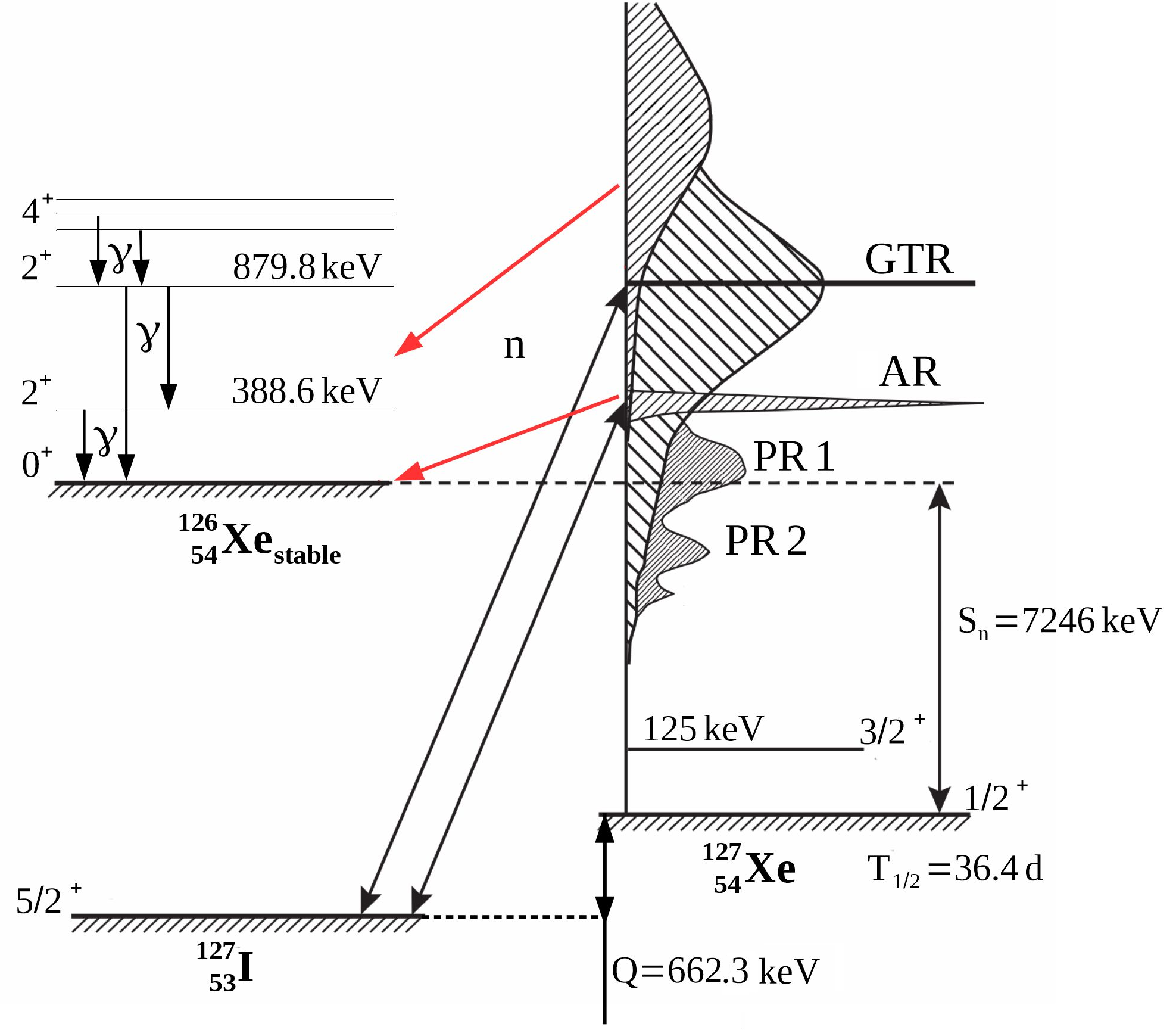}
\caption{Scheme  of  charge-exchange  excitations of the $^{127}$Xe nucleus in the $^{127}$I$(p,n) ^{127}$Xe reaction with the decay of high-lying excitations in the stable $^{126}$Xe isotope  accompanied  by  the  emission  of  a  neutron.  The giant Gamow–Teller resonance (GTR), analog resonance(AR),  and  lower  lying Gamow–Teller pygmy  resonances  (PR)  are  indicated; $S_n$ is the neutron separation energy in the $^{127}$Xe nucleus}
\label{fig:1}
\end{figure}

Fig. \ref{fig:1} shows the scheme of excited states of the $^{127}$Xe isobaric nucleus, various regions of the excitation spectrum, and isotopes formed as a result of neutrino capture by the $^{127}$I nucleus and subsequent decay. 
The respective experimental data were obtained in the reaction $^{127}$I$(p,n) ^{127}$Xe \cite{Palarczyk_PhysRevC.59.500} and a table of values of matrix elements $B(GT)$ depending on the energy $E_x$ (with 0.5~MeV step) in the daughter nucleus $^{127}$Xe up to the energy 20~MeV was presented.
It was found that the total sum of $B(GT)$ up to 20~MeV energy is $53.54 \pm 0.22$ \cite{Palarczyk_PhysRevC.59.500}, which is $\approx 85\%$ of total GT summ rule $3(N - Z) = 63$ for $^{127}$I. Below, we will discuss the reasons behind this underestimation (\emph{quenching}-effect).

There are also shown in Fig. \ref{fig:1}: threshold energy $Q_{\beta} = 662.3 \pm 2.0$~keV and  $S_n = 7246 \pm 5$~keV - neutron separation energy in  $^{127}$Xe nucleus \cite{nds.iaea.org}.
Excited states of the $^{127}$Xe nucleus at energies above $S_n$ will  decay  with  neutron  emission  to  the  stable  $^{126}$Xe isotope;  as  a  result,  neutrino  capture  on  the  $^{127}$I nucleus in reaction (\ref{eq:127I_capture}) will lead to the formation of the $^{127}$Xe  and $^{126}$Xe  isotopes \cite{LutostanskyEtAl:2020_JETP_lett}.
If the daughter $^{126}$Xe nucleus is born on the excited state, the subsequent deexcitation will lead to the emission of one or several gamma-ray  photons.
The yield of the stable $^{126}$Xe isotope is relatively low, but it survives in the xenon fraction after long-term storage following the decay of the $^{127}$Xe ($T_{1/2} = 36.4$~d) isotope.
In addition, according to Fig. \ref{fig:1}, the production of the $^{126}$Xe isotope is accompanied by a transition from the low-lying excited   $2^+$   state $^{126}$Xe to the ground state   $0^+$   with the  emission  of  gamma-ray  photons  with  the  energy $E_1=388.6$~keV.
Thus,  the  analysis  of  the  $^{126}$Xe/$^{127}$Xe isotope ratio in the formed xenon gas mixture and the detection of gamma emission in $^{126}$Xe open new capabilities of the iodine detector for the detection of solar neutrinos and make it possible to separate the important boron component of the solar spectrum.

\begin{figure}[ht!]
\centering
\includegraphics[width=0.7\linewidth]{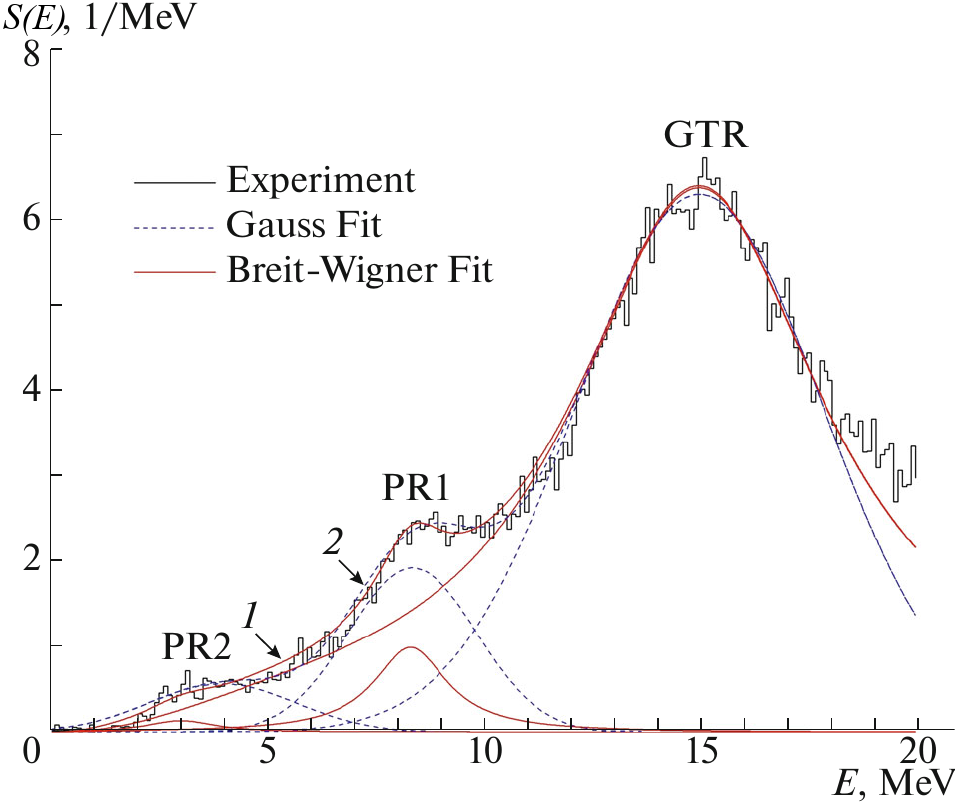}
\caption{Charge-exchange strength function $S(E)$ distribution for the $^{127}$Xe nucleus according to the analysis of  $^{127}$I$(p,n) ^{127}$Xe reaction experimental data \cite{Palarczyk_PhysRevC.59.500}.  Giant Gamow–Teller resonance (GTR) and two pygmy resonances PR1 and PR2 were fitted by Gaussian (G) or by Breit–Wigner (B-W) shape forms. The summed dependencies $S$(E)=$S$(GTR) + $S$(PR1) +$S$(PR2) are given in two approximations: (\emph{1}) B–W and (\emph{2}) G.}
\label{fig:2}
\end{figure}

Fig. \ref{fig:2} represents charge-exchange strength function $S(E)$  for the $^{127}$Xe using experimental data analysis of  $^{127}$I$(p,n) ^{127}$Xe reaction \cite{Palarczyk_PhysRevC.59.500}. 
$S(E)$ could be submitted  as sum of strength functions of three resonances: Giant Gamow–Teller resonance $S(\mathrm{GTR})$ and two pygmy ones $S(\mathrm{PR1})$, $S(\mathrm{PR2})$.

Resonances  were fitted by Gaussian (G) or by Breit–Wigner (B-W) shape forms. 
The summed dependances $S(E) = S(\mathrm{GTR}) +
S(\mathrm{PR1}) + S(\mathrm{PR2})$ in the Fig. \ref{fig:2} are given for two approximations: (\emph{1}) B–W and (\emph{2}) G.

GTR peak energy $E_{\mathrm{GTR}} =14.9$~MeV is the same for both fits  as well as PR1 peak energy $E_{\mathrm{\mathrm{PR1}}} = 8.3$~МeV.
Experimental values from \cite{Palarczyk_PhysRevC.59.500} are closer to B-W fit results: $E_{\mathrm{GTR}} = 14.5$~МeV, $E_{\mathrm{PR1}} \approx 8.7$~МeV, $E_{\mathrm{PR2}} = 5-6$~МeV and $E_{\mathrm{PR3}} = 3.08$~МeV.

\section{Strength function calculation}

Within the microscopic theory of finite Fermi systems  (TFFS),  charge-exchange  excitations  of  nuclei are described by the system of equations for the effective field in the form \cite{Migdal_book}:
\begin{equation} 
\label{eq:V_pn,V_ph}
    \begin{aligned}
         V_{pn} = e_q V^{\omega}_{pn} + \sum_{n^{'}p^{'}} 
         F^{\omega}_{np,n^{'}p^{'}} \rho_{p^{'}n^{'}}, \mkern 30mu
         V^{h}_{pn} =  \sum_{p^{'}n^{'}} F^{\omega}_{np,n^{'}p^{'}} \rho^{h}_{p^{'}n^{'}}
    \end{aligned}
\end{equation}
where $V_{pn}$ and $V^{h}_{pn}$ are the effective fields of quasiparticles and holes in the nucleus, respectively, and $V^{\omega}_{pn}$  is the external charge-exchange field.
The system of secular equations (\ref{eq:V_pn,V_ph}) is solved for allowed transitions with a  local  nucleon–nucleon  interaction  $F_{\omega}$  in  the  Landau–Migdal form \cite{Migdal_book}
\begin{equation} 
\label{eq:F_omega}
    F_{\omega} = C_{0}(f^{'}_{0} + g^{'}_{0} (\vec{\sigma_{1}} \vec{\sigma_{2}}))(\vec{\tau_{1}}\vec{\tau_{2}})\delta(\vec{r_{1}} - \vec{r_{2}})
\end{equation}
where $C_{0} = (\mathrm{d} \rho / \mathrm{d} \epsilon_{F})^{-1} = 300~$MeV$\cdot$\emph{fm}$^{3}$ ($\rho$~is the average nuclear-matter density) and $f^{'}_{0}$  and $g^{'}_{0}$ are the parameters of isospin–isospin and spin–isospin quasiparticle interactions, respectively.
These coupling constants are phenomenological and could be determined  by fitting experimental data.

The inclusion of terms associated with the pion mode leads to an effective renormalization of the coupling constant $g^{'}_{0}$ \cite{Lutostansky2011_PhysicsOfAtomicNuclei}
\begin{equation} 
\label{eq: g'_0eff}
    \begin{aligned}
        g^{'}_{0eff} = g^{'}_{0} - \Delta g^{'}_{\pi} \\
    \end{aligned}
\end{equation}
where $\Delta g^{'}_{\pi}$ is the correction to $g^{'}_{0}$ for the effect of the pion mode associated primarily with a high-lying $\Delta$-isobar.
According to the calculations performed in \cite{Lutostansky2011_PhysicsOfAtomicNuclei} with considering the pion mode, these effects exert an influence on states lying substantially higher than GTR.
The values of $f^{'}_{0} = 1.35$ and $g^{'}_{0} = 1.22$ were obtained earlier in \cite{Lutostansky2011_PhysicsOfAtomicNuclei} from a comparison of the calculated and measured values of the GTR and AR energies.
However, the recent analysis \cite{Lutostansky2020} of calculated and experimental data on the energies of analog (for 38 nuclei) and Gamow–Teller (for 20 nuclei) resonances showed that the local-interaction parameters should be slightly corrected \cite{Lutostansky2020}:

\begin{equation}
\label{eq:g'f'}
    \begin{aligned}
    f^{'}_{0} = 1.351 \pm 0.027,  \mkern  40mu  g^{'}_{0} = 1.214 \pm 0.048 
    \end{aligned}
\end{equation}

The calculations of charge-exchange excitations for the $^{127}$I isotope were performed with the inclusion of this correction.
The energies, $E_i$, and the squares of the matrix elements, $M^{2}_{i}$ formed by allowed transitions for excited isobaric states of the $^{127}$Хе daughter nucleus were calculated. 
The matrix elements are normalized according to the sum rule for GT transitions as in \cite{Lutostansky_Shulgina_PhysRevLett.67.430}:
\begin{equation} 
\label{eq:Summ.M_i}
    \sum M^{2}_{i} = q[3(N-Z)] = e^{2}_{q}[3(N-Z)] \approx \int_{0}^{E_{max}} S(E) dE = I(E_{max})
\end{equation}
where $q < 1$ is the parameter determining the \emph{quenching}-effect (deficit in the sum rule to the maximum theoretical value $3(N-Z)$ at $q = 1$ \cite{ARIMA1999260}). Within the TFFS framework, $q=e^{2}_{q}$,  where  $e_q$  is  the  effective  charge  \cite{Migdal_book}; $S(E)$  is the charge-exchange strength function and $E_{max}$ is the maximum  energy  taken  into  account  in  the  calculations or in the experiment.
In the present calculations we used the value $E_{max} = 20$~MeV, as in the experiment reported in \cite{Palarczyk_PhysRevC.59.500}.
The value $q$~=~0.85 was obtained in \cite{Palarczyk_PhysRevC.59.500} for $^{127}$I, which is close to the calculated one. Should be noted that for other nuclei, the calculated values of $e_q$ differ from 0.9, mainly to a lower side \cite{Lutostansky_Tikhonov2018} \cite{Lutostansky:2019iri}. 
In experiments, the sum in Eq. (\ref{eq:Summ.M_i}) rarely reaches $q$~=~0.70 (70$\%$) of $3(N-Z)$, primarily because of moderately small values of the energy $E_{max}$, difficulties encountered in separating and subtracting backgrounds at energies above $E_{\mathrm{GTR}}$ as well as the result of two-particle-two-hole excitations or $\Delta$–isobar-neutron-hole excitation.

Since the spectrum of the function $S(E)$ has a continuous resonance structure, the Breit–Wigner form was chosen for partial amplitudes $S_{i}(E)$: 
\begin{equation} 
\label{eq:Si(E)_form}
    \begin{aligned}
        S_{i}(E) = M^{2}_{i} \frac{\Gamma_i}{(E-E_i)^2+\Gamma^{2}_{i}} \phi (E)
    \end{aligned}
\end{equation}
where $\phi(E) = 1 - e^{-(\frac{E}{\Gamma_{i}})^2}$ is shape factor \cite{Migdal_book} responsible for equality $S(E) = 0$ (or $\phi(E)=0$) at energies $E \leqslant Q$ i.e. energies lower than reaction threshold.

According to \cite{Migdal_book} the width was chosen in the form
\begin{equation} 
\label{eq:width_as_sum}
    \Gamma(E) = \alpha E^2 + \beta E^3 + ...
\end{equation}
In calculating $S(E)$ it suffices to use only the first term of the series with $\alpha \approx \epsilon^{-1}_F$, which effectively takes into account three-quasiparticle configurations.
We used the value $\alpha= 0.018$~MeV$^{-1}$, obtained from the averaged experimental widths of GTR resonances.
The low-lying discrete states were performed separately.

\section{Energies of resonances and charge-exchange strength function for the $^{127}$Xe isotope}
\begin{figure}[ht!]
\centering
\includegraphics[width=0.7\linewidth]{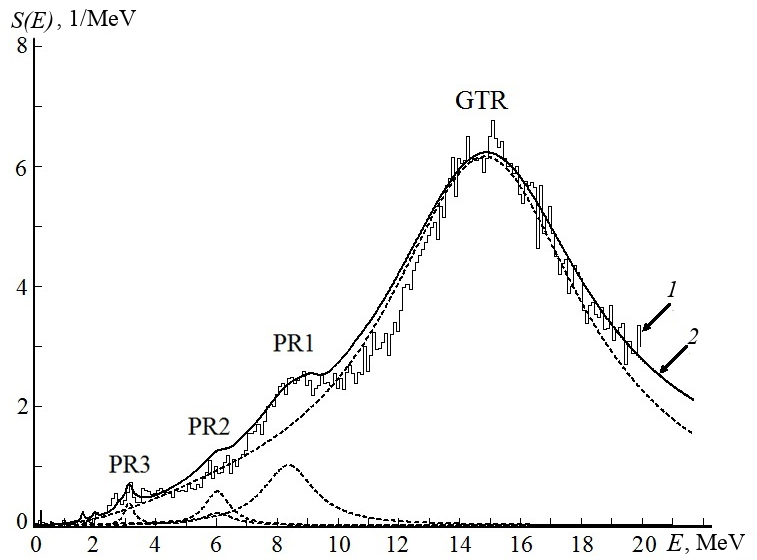}
\caption{Charge-exchange  strength  function  $S(E)$  of  the $^{127}$Xe isotope for GT excitations of $^{127}$I. Solid line \emph{1} shows experimental  data  on  the  $^{127}$I$(p,n) ^{127}$Xe  reaction  from \cite{Palarczyk_PhysRevC.59.500}, solid line \emph{2} is our TFFS calculation, and the dashed lines correspond to GTR, PR1, PR2, and PR3 resonances.}
\label{fig:3}
\end{figure}

The calculated charge-exchange strength function $S(E)$ for the $^{127}$Xe isotope is shown in Fig. \ref{fig:3} together with experimental data on the $^{127}$I$(p,n) ^{127}$Xe reaction \cite{Palarczyk_PhysRevC.59.500}.
The strength function $S(E)$ has a resonance character associated with collective excitations of the parent nucleus that receive contributions from various quasiparticle $(p\bar{n})$-configurations.
Of them, dominant ones are determined by $\Delta j = - j_+ - j_-(n-p)$ quasiparticle spin–orbit transitions (spin-flip transitions - \emph{sft}), such as the $1h_{11/2} - 1h_{9/2} (h)$, $2d_{5/2} - 2d_{3/2} (d)$, $1g_{9/2} - 1g_{7/2} (g)$ and to a less extent, by $\Delta j = 0 = j -j$ (core polarization states - \emph{cps}) transitions  such as the $1h_{11/2} - 1h_{11/2} (h)$, $2d_{3/2} - 2d_{3/2} (d)$, $1g_{7/2} - 1g_{7/2}$ and $3s_{1/2} - 3s_{1/2}$ transitions.
Back-spin-flip quasiparticle excitations with $\Delta j = +1 = j_- - j_+$ (back-spin-flip states — \emph{bsfs}) also contribute: these are $2d_{3/2} - 2d_{5/2}$ and $1g_{7/2} - 1g_{9/2}$ transitions.

\begin{table}[ht]
\caption{Contribution (in \%) of single-particle $(n-p)$ transitions to the structure of charge-exchange excitations of the $^{127}$Хе nucleus}
\centering
\begin{adjustbox}{width=0.9\textwidth}
\small
\begin{tabular}{| c | c | c | c | c | c | c | c |} 
    \hline
    \multirow{3}{*}{\makecell{ Excit. \\ type}} & \multicolumn{2}{|c|}{$E$,~MeV} & \multicolumn{5}{|c|}{Contribution to structure of excitations, $\%$} \\ 
    \cline{2-8}
     & \makecell{Exper. \\ \cite{Palarczyk_PhysRevC.59.500}} & \makecell{Calc. \\ TFFS} & $1h_{11/2} - 1h_{9/2}$ & $2d_{5/2} - 2d_{3/2}$ & $1g_{9/2} - 1g_{7/2}$ & $j - j$ & $j_- - j_+$ \\
     \hline
     GTR & 14.5 & 14.6 & 29 & 12 & 44 & 12 & 3 \\ 
     
      
     PR1 & 8.4 & 8.3 & - & 41 & 33 & 22 & 4 \\
     
     PR2 & 5.5 - 6.5 & 6.3 & - & - & - & 94 & 6 \\

     PR3 & 3.08 & 3.1 & - & 21 & 6 & 68 & 5 \\
     
      & 2.62 & 2.8 & - & - & - & 96 & 4 \\
      
      & & 2.0 & - & - & - & 13 & 87 \\
    \hline
\end{tabular}
\end{adjustbox}
\label{table:2}
\end{table}

Table \ref{table:2} presents the contributions of single-particle $(n-p)$ transitions into the structure of charge-exchange excitations of $^{127}$Xe nucleus both according to experimental data from \cite{Palarczyk_PhysRevC.59.500} in the reaction $^{127}$I$(p,n) ^{127}$Xe and according to the results of TFFS calculations.
The Gamow–Teller resonance (GTR) at 14.6~MeV (the experiment in \cite{Palarczyk_PhysRevC.59.500} yields $E_{\mathrm{GTR}}$=14.5~MeV) is the most collective state.
Quasiparticle transitions for which $\Delta j = -1 = j_+ - j_-$ make a dominant contribution to the GTR structure (in all, 85$\%$); an excitation formed primarily by spin-flip transitions of the $h$ type lies lower.
The calculated PR1 pygmy resonance is close to the respective experimental value; a dominant contribution to its structure comes from spin-flip transitions of the $d$ and $g$ types, and \emph{cps} transitions of the $j-j$ type also contribute (at a level of 22\%).
The PR2 pygmy resonance was not found experimentally; theoretically, it is interpreted in terms of two $\Delta j$ = 0 \emph{cps} excitations belonging to the $(j-j)$ type.
This means that, according to the terminology adopted in \cite{Lutostansky2011_PhysicsOfAtomicNuclei}, this is a split collective excitation of the $\omega_0$ type.
The PR3 resonance is determined by spin-flip transitions of the $d$ type and, primarily, by $\Delta j$ = 0 \emph{cps} transitions.
According to calculations, there are two single-particle states at energy range between 2 and 3~MeV: a state at 2.8~MeV determined by $j-j$ transitions of the \emph{cps} type [that is, $1h_{11/2} - 1h_{11/2}$] and a state at 2.0~MeV determined by \emph{bsfs} transitions [$2d_{3/2} - 2d_{5/2}$].
States lying below 2~MeV are primarily single-particle states, and we do not consider them in the present study.

\begin{figure}[ht!]
\centering
\includegraphics[width=0.7\linewidth]{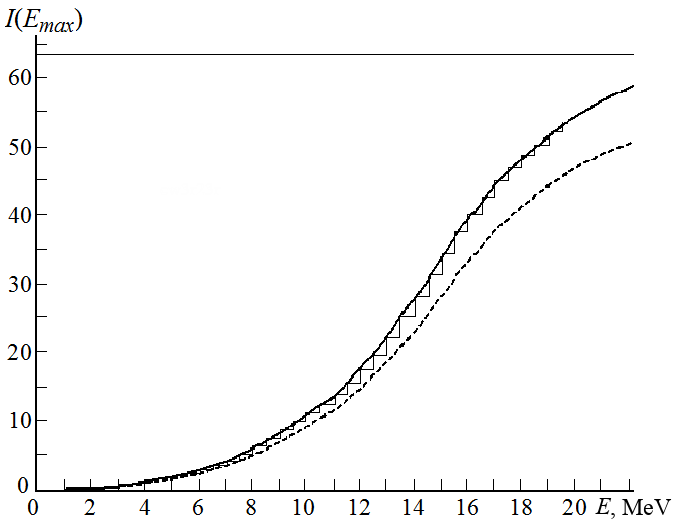}
\caption{Integral $I(E_{max}$ given by Eq. (\ref{eq:Summ.M_i}) versus the energy $E_{max}$ for the $^{127}$Xe isotope: (steps) experimental data from \cite{Palarczyk_PhysRevC.59.500},  (solid  line)  calculation  with  $e_q  =  0.9$,  (dashed  line) calculation  with  $e_q  =  0.8$,  and  (horizontal  straight  line) sum-rule value of $3(N-Z) = 63$.}
\label{fig:4}
\end{figure}

Of particular interest are the problem of the sum rule in Eq. (\ref{eq:Summ.M_i}) and the related \emph{quenching}-effect consisting in the shortage of the sum in Eq. (\ref{eq:Summ.M_i}) with respect to the maximum theoretical value of $3(N - Z)$ \cite{ARIMA1999260} at $q$ = 1.
The experimental value of the quenching parameter may change greatly from one nucleus to another  \cite{Lutostansky_Tikhonov2018} \cite{Lutostansky:2019iri} - for example, from $q$ = 0.67 or $67 \pm 8\%$ for $^{98}$Mo \cite{Rapaport_98Mo_PhysRevLett.54.2325} to $q$ = 0.85 or 85\% in the case of the $^{127}$I nucleus \cite{Palarczyk_PhysRevC.59.500}.

The  integral  $I(E_{max})$  for  the  isotope  $^{127}$Xe given by Eq. (\ref{eq:Summ.M_i}) is shown in Fig. \ref{fig:4} as a function of the energy $E_{max}$.
It is seen that the best description of the experimental data in this case is provided by the calculations with the effective-charge value of $e_q  =  0.9$ ($q = 0.81$).
The  experiment  with  $^{127}$I  reported  in  \cite{Palarczyk_PhysRevC.59.500}  gives  the value $q = 0.85$, close to the calculated value.
It is noteworthy  that  the  calculated  $e_q$ values  for  other  nuclei differ from 0.9, predominantly on the side of smaller values \cite{Lutostansky_Tikhonov2018} \cite{Lutostansky:2019iri}.
Mostly, this is characteristic of nuclei lighter  than  $^{127}$I  and  is  due  partly  to  the  disregard  of high-lying (above GTR) excitations in the experiment that are formed by single-particle transitions in which $\Delta n = 1, 2$.

\section{Neutrino capture cross-section by the  $^{127}$I nuclei and Fermi-function}

The $(\nu_e, e^-)$ cross-section, which depends on the incident-neutrino energy $E_{\nu}$, has the form \cite{Lutostansky_Shulgina_PhysRevLett.67.430}:
\begin{equation} 
\label{eq:sigma(E)}
    \begin{aligned}
        \sigma(E_{\nu}) = \frac{(G_F g_A)^2}{\pi c^3 \hbar^4} \int_{0}^{W - Q} W p_e F(Z, A, W) S(x) dx \\
        W = E_{\nu} - Q - x - m_e c^2 \\
        cp_e = \sqrt{W^{2} - (m_e c^2)^2}
    \end{aligned}
\end{equation}
where $F(Z, A, W)$ is the Fermi-function, $S(E)$ is the charge-exchange strength function, $G_F / (\hbar c)^3 = 1.1663787(6) \times 10^{-5}$~GeV$^{-2}$ is the weak coupling constant and $g_A = - 1.2723(23)$ is the axial-vector constant from \cite{PDG_2020}.

Since the change in the Fermi-function according to the Eq. (\ref{eq:sigma(E)})  is practically proportional to the change in the cross-section it is essential to allude  different approaches how it could be calculated. For more details one could look  \cite{fermi_function_2017}.

General expression for $F(Z, A, W)$ firstly was done into Fermi's paper \cite{Fermi:1934hr}:
\begin{equation}
    F_0(Z, A, W) = 4(2pR)^{2(\gamma -1)}  \frac{|\Gamma(\gamma + iy)|^2}{(\Gamma(1+2\gamma))^2}e^{\pi y}, \gamma = \sqrt{1-(\alpha Z)^2} ,     y = \pm \alpha ZW/p 
    \label{eq:fermi_func_F_0}
\end{equation}

Finite nucleus size can be described by the factor $ L_0 $, such that:
$$ F(Z, A, W) = F_{0} \cdot L_{0}. $$ Sufficiently exact expression for $ L_0 $ is model dependent. 
There is a wide use of the Fermi function in which $ L_0 $ is obtained by a numerical solution of the Dirac equation \cite{Janecki_1969}:

\begin{equation}
    \label{eq:Janecke}
    L_0 = 1 \mp \frac{13}{15} \alpha ZWR + \dots
\end{equation}

More extensive calculations made by Wilkinson \cite{WILKINSON1990509} gives us next expression for $L_0$:

\begin{equation}
\label{eq:L_0}
    L_0 = 1 +  ~\frac{13}{60} (\alpha Z)^2  \mp  \frac{\alpha ZWR(41-26\gamma)}{[15(2\gamma -1)]}  \mp  \frac{\alpha Z R \gamma (17-2\gamma)}{[30 W (2\gamma-1)]}  +  \Omega
\end{equation}

\begin{figure}[ht!]
\centering
\includegraphics[width=0.7\linewidth]{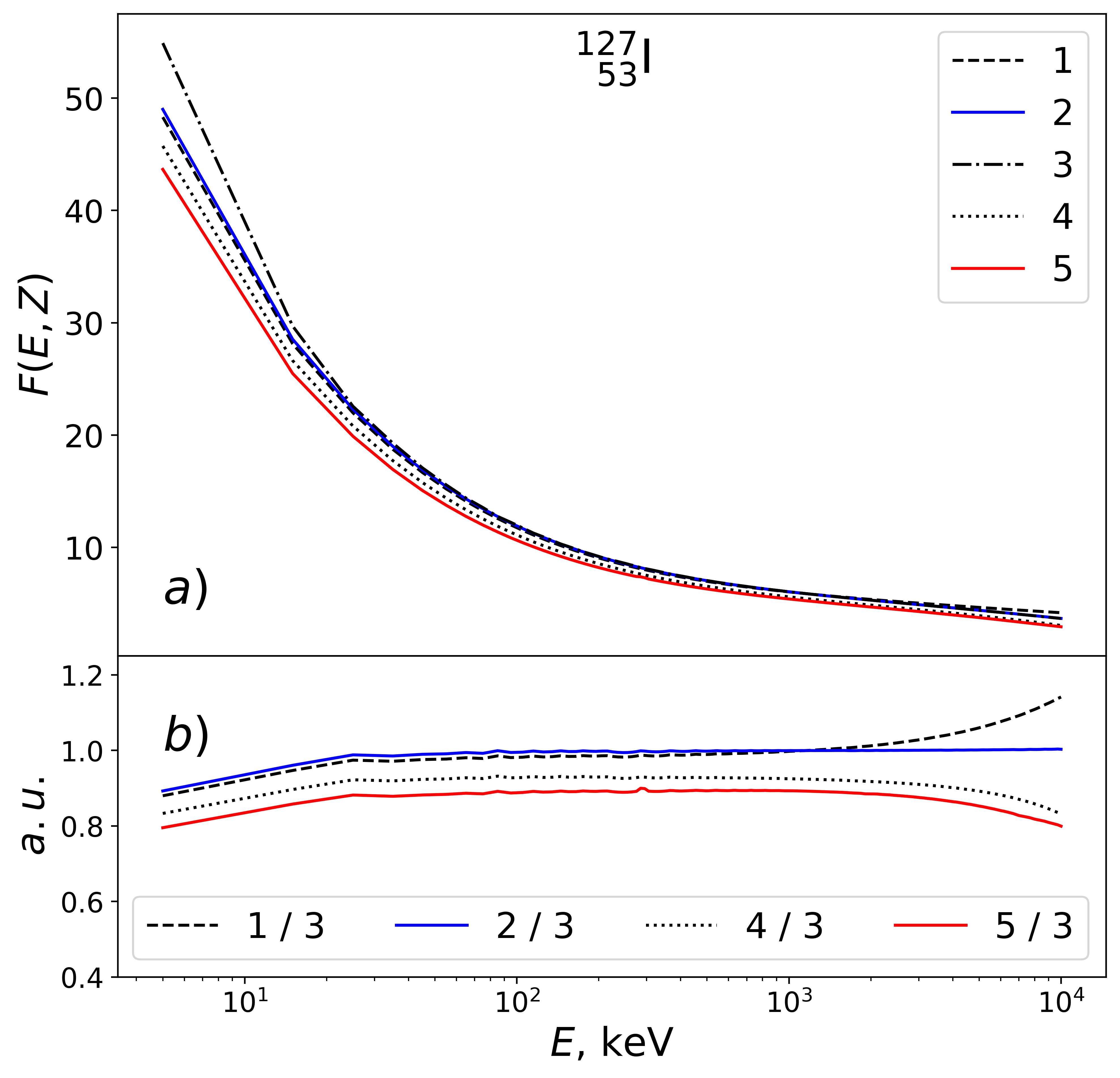}
\caption{\emph{a)} -- Fermi-function calculations: 1 - \cite{Fermi:1934hr}, 2 -  \cite{fermi_function_2017} with $L_0$ and spatial charge distribution 
 factors  , 3 - \cite{Janecki_1969}, 4 - \cite{DZHELEPOV_ZYRIANOVA_1956}, 5 - \cite{Suslov_1968}. \emph{b)} Fermi-functions normalized on \cite{Janecki_1969}. Numbers are the same like in \emph{a)}}
\label{fig:fermi_func_comparison}
\end{figure}

The correction term $\Omega$ is presented in \cite{fermi_function_2017}.
Papers \cite{DZHELEPOV_ZYRIANOVA_1956}, \cite{Suslov_1968} have importance for  nuclei screening estimations and its influence on Fermi-function computation. Fig. \ref{fig:fermi_func_comparison} demonstrates comparison of few variants of Fermi-functions. It is seen that the greatest differences (up to 15$\%$) are observed at high energies.

The question how to describe and measure the nuclear radius is highly nontrivial. For medium mass isotopes ($A \approx 100$) it is conventional to apply $R = r_0 \cdot A^{1/3} = 1.20 \cdot A^{1/3}$~\emph{fm} formula \cite{Bohr_Motteleson_book}.

Spatial charge distribution $\rho(r)$ could change expression for $R$.
So, departing from uniform distribution, and assuming 
\begin{equation}
\label{eq:rho(r)}
    \rho(r) = \rho_0 \cdot [1+\exp[(r-R)/a]]^{-1}
\end{equation}
if $\rho_0 = 0.17$ $\emph{nucl.} \cdot \emph{fm}^{-3}$ и $a = 0.54~fm$, we will have \cite{Bohr_Motteleson_book}:
\begin{equation}
\label{eq:rho(r)}
    R \approx (1.12 \cdot A^{1/3} - 0.86 \cdot A^{-1/3} + \dots)~\emph{fm}
\end{equation}

More information and an improved $ R_C(A) $ dependence and  charge distributions  for $R$ could be obtained from the analysis of data on analog resonances as in the works \cite{Lutostansky_Tikhonov_2015_nuclear_size}, \cite{Anderson_1965}, \cite{BATTY1966}. 
Thus, it was obtained for the charge radius (for nuclei with $ A \ge 40 $) $ R_C = 1.25 \cdot A^{1/3}~\emph{fm} $.

Another relation between $R$ and Fermi-function founded by \cite{Elton_nuclear_sizes_1961} is presented in Bahcall's book (Eq. (8.16) from \cite{Bahcall_book}):

\begin{equation}
\label{eq:R_bahcall}
    R = (2.908 \cdot A^{1/3} + 6.091 \cdot  A^{-1/3} - 5.361 \cdot A^{-1}) \cdot 10^{-3} (\hbar/m_ec)
\end{equation}

After averaging the Fermi-function over the nuclear volume by the radius $ R $, a small correction for the average value (Eq. (8.17) from \cite{Bahcall_book}) will be : 

\begin{equation}
\label{eq:fermi_bahcall}
    <F(Z, E)> = [1 -  \frac{2}{3} (1 - \gamma)]^{-1}  \cdot F(Z, E) 
\end{equation}

Calculation using this Eq. (\ref{eq:fermi_bahcall}) gives a fairly good approximation for the Fermi-function.

Let us note the calculations of the charge radii in the microscopic approach in the framework of the TFFC theory with the Fayans density functional \cite{Saperstein2016}.
This calculations were carried out mainly for neutron-deficient spherical nuclei, and the best accuracy was shown for microscopic approaches.
For such nuclei, the deviations from the dependence $ R \sim A^{1/3} $ are relatively large and approach it near the stability line.

\begin{figure}[ht!]
\centering
\includegraphics[width=0.7\linewidth]{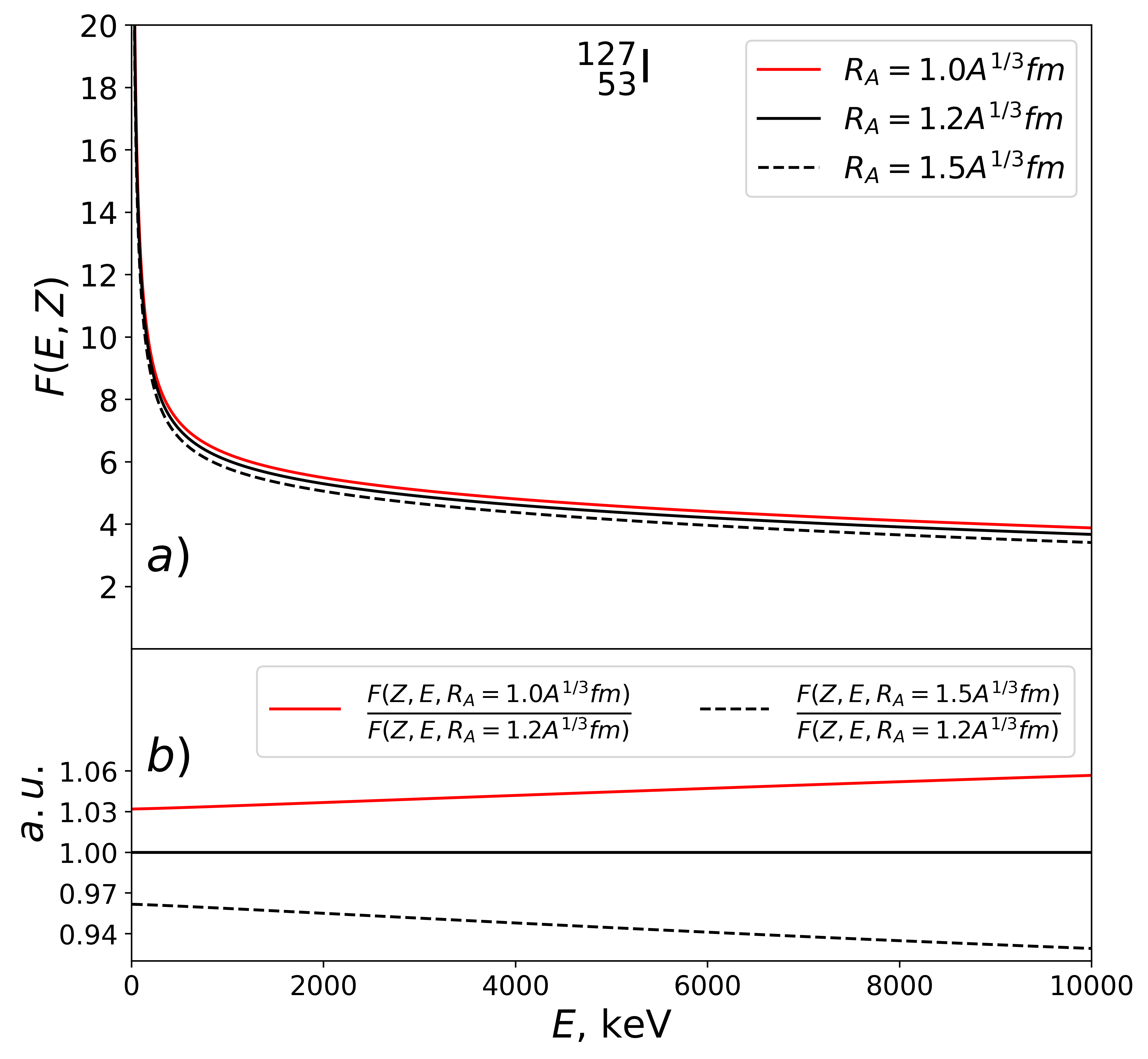}

\caption{\emph{a)} Fermi-function \cite{fermi_function_2017} with 
$L_0$ and spatial charge distribution factors calculated for three 
values of the parameter $r_{0}$ in $R = r_{0} \cdot A^{1/3}$: $r_0 = 1.0$,  $r_{0} = 1.2$ and $r_{0} = 1.5$. \emph{b)} -  their ratios.}
\label{fig:fermi_func_comparison_R}
\end{figure}

The isotopic dependence of the charge radii in a long chain of copper isotopes \cite{Borzov2020} was recently investigated.
The same approach was used as in the work \cite{Saperstein2016}: the self-consistent TFFC with the Fayans density functional.
Systematic deviation from the $ R \sim A^{1/3} $ dependence was observed towards a decrease with increasing neutron excess.
Analysis of the experimental data used in \cite{Borzov2020} showed that for neutron-deficient copper isotopes the deviations from the dependence $R = 0.96 \cdot A^ {1/3}$~\emph{fm} are $\delta R \approx -2.2\%$ ($^{58}$Cu), and for neutron-abundant nuclei $\delta R \approx +3.1\%$ ($^{78}$Cu).
Note that in the region of stable nuclei the deviations are insignificant ($\delta R \approx 1.0\%$).
Even-odd oscillations are also observed, which depend on the shell structure.

Calculations using the density functional were carried out recently for chains of potassium isotopes ($^{36-52}$K, $Z=19$)  and more heavier nuclei Ca, Sn, Pb \cite{Reinhard}. So for isotopes $^{36-52}$K \cite{Koszorus:2020mgn} the fitting gives deviations from the dependence  $ R = 0.98 \cdot A^{1/3} \emph{fm},~\delta R \le \pm 4.0 \% $, for $^{108-134}$Sn \cite{tin_isotopes} isotopes $ R = 0.94 \cdot A^{1/3} \emph{fm},~\delta R \le \pm 2.0 \% $.

How important is the dependence of the Fermi-function on the radius can be seen from Fig. \ref{fig:fermi_func_comparison_R}.
So with an increase in the parameter $ r_{0} $ by 25 $\%$ from 1.2 to 1.5, the value of the Fermi-function decreases linearly by 4 $\%$ at energy $ E = 10 $ ~ MeV.
For the $ ^{127}$I isotope itself, theoretical estimates of the nucleus parameters still require special calculations; nevertheless, the characteristic values for heavy nuclei can be estimated as in  \cite{fermi_function_2017}.

\section{Effect of charge-exchange resonances on the neutrino capture cross-section of the $^{127}$I nucleus}

\begin{figure}[ht!]
\centering
\includegraphics[width=0.7\linewidth]{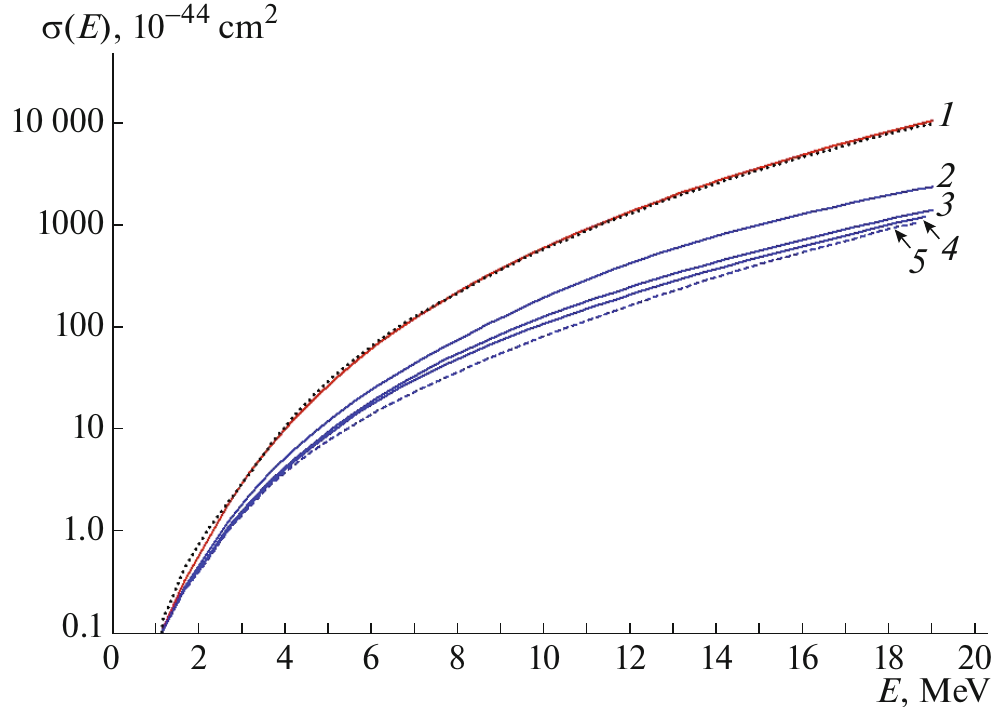}

\caption{Neutrino-capture cross-section  $\sigma(E)$in the reaction $^{127}$I$(\nu_e,e^-) ^{127}$Xe.  The points on display stand for the results of the calculations with the experimental strength function $S(E)$ (see Fig. \ref{fig:2}). The  solid and dashed curves represent the results of the calculations performed with the strength function $S(E)$ obtained within the TFFS approach (see Fig. \ref{fig:3}):  $(\emph{1})$ summed cross-section; $(\emph{2})$ cross-section calculated without GTR; $(\emph{3})$ cross-section calculated without GTR and PR1; $(\emph{4})$ cross-section calculated without GTR, PR1, and PR2; and $(\emph{5})$ cross-section calculated without GTR, PR1, PR2,and PR3}
\label{fig:cross-sec_1}
\end{figure}

\begin{figure}[ht!]
\centering
\includegraphics[width=0.7\linewidth]{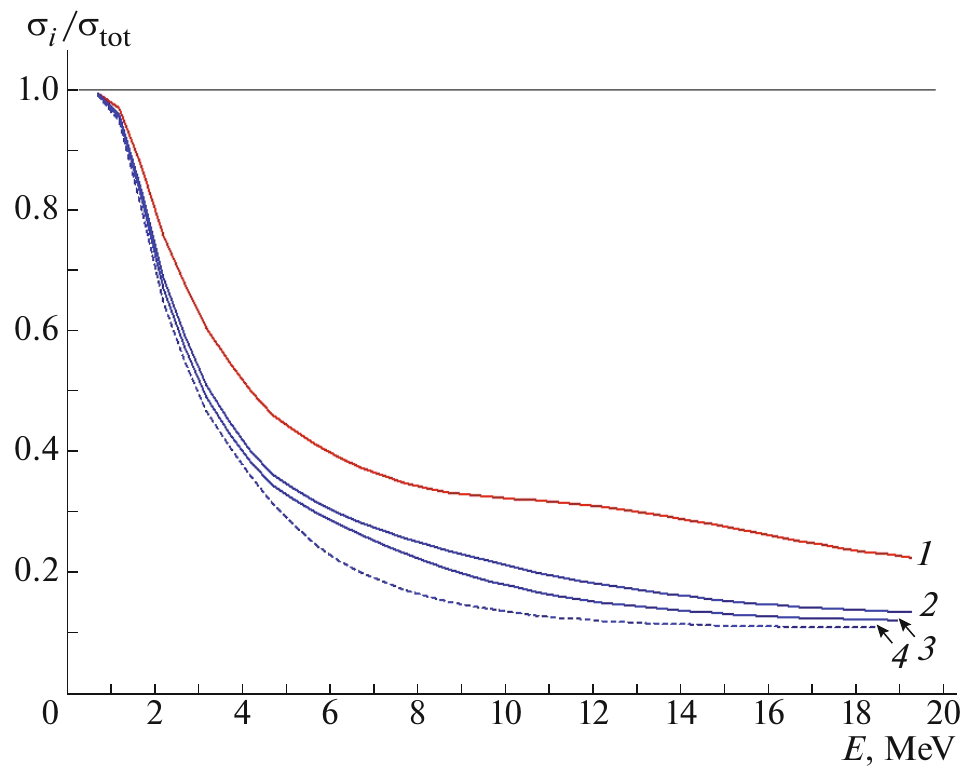}
\caption{Ratios  of  the  cross-sections $\sigma_i(E)$ calculated  for  the  reaction $^{127}$I$(\nu_e,e^-) ^{127}$Xe  and  normalized  to the summed cross-section $\sigma_{tot}(E)$ determined within the TFFS approach:  $(\emph{1})$ results calculated without GTR; $(\emph{2})$ results calculated without GTR  and  PR1;  $(\emph{3})$ results calculated without GTR, PR1,  and PR2;  and $(\emph{4})$ results calculated without GTR, PR1, PR2, and PR3.}
\label{fig:cross-sec_2}
\end{figure}

The cross-sections $\sigma(E)$ for neutrino capture by $^{127}$I nuclei in the reaction $^{127}$I$(\nu_e,e^-) ^{127}$Xe (Fig.~\ref{fig:cross-sec_1} and Fig.~\ref{fig:cross-sec_2}) were calculated with the experimental charge-exchange strength functions $S(E)$ (see Figs.~\ref{fig:2} and \ref{fig:3}) and with the strength functions $S(E)$ (see Fig.~\ref{fig:3})calculated within the TFFS framework according to the method used in \cite{Lutostansky_Tikhonov2018}.
In order to analyze the effect of charge-exchange resonances on the cross-section $\sigma(E)$, we have also performed calculations without taking into account GTR and  pygmy resonances. 

Maximum deviation of 30\% to 15\% from the cross-section summed over all resonances is observed at energies between 1.5 and 2.5~MeV.
At energies above 6~MeV, the discrepancies do not exceed 10\%.
The disregard of only two resonances, GTR and PR1, reduces the cross-section  by about 25\% to about 80\% as the neutrino energy changes from 2 to 12~MeV.
It can be seen more clearly in Fig. \ref{fig:cross-sec_2}, which shows the ratios of  the  cross-sections $\sigma_i(E)$ calculated  for  the  reaction $^{127}$I$(\nu_e,e^-) ^{127}$Xe  and  normalized  to the summed cross-section $\sigma_{tot}(E)$ determined within the TFFS approach. Contribution from AR resonance  to $\sigma(E)$  is negligible (less than 1\%) for the reaction $^{127}$I$(\nu_e,e^-) ^{127}$Xe.

\section{Solar neutrino capture rate by the $^{127}$I nucleus}

In  order  to  calculate  capture  cross-sections for solar  neutrinos  and  to  analyze  the  effect  of  charge-exchange resonances, it is important to simulate correctly the flux of solar neutrinos.
At the present time, there are many models of the Sun.
They differ from one  another  in  the  relative  concentration  of  helium and elements heavier than helium (metallicity) or may differ in the concentration of some specific element in some part of the Sun (at the center or at the surface).
Also,  the size of the Sun’s convective zone and parameters of medium nontransparency may be different in  different  models.
The  BS05(OP),  BS05(AGS, OP),  and  BS05(AGS,  OPAL)  models  developed  by the group headed by Bahcall \cite{Bahcall_2005} are the most popular models at the present time.
There are newer models as  well  developed  by  another  group.
They  include the  B16-GS98  and  B16-AGSS09met  models  \cite{Serenelli_2017} and other.
Although  B16  models  are  newer  than  BS05(OP)models  and  also  agree fairly well  with  observational data, we present here only BS05(OP) data.
Since all known  models  differ  only  in  the  normalization  of  neutrino fluxes from each nuclear reaction in the Sun, one can readily rescale the data to another model.

\begin{figure}[ht!]
\centering
\includegraphics[width=0.8\linewidth]{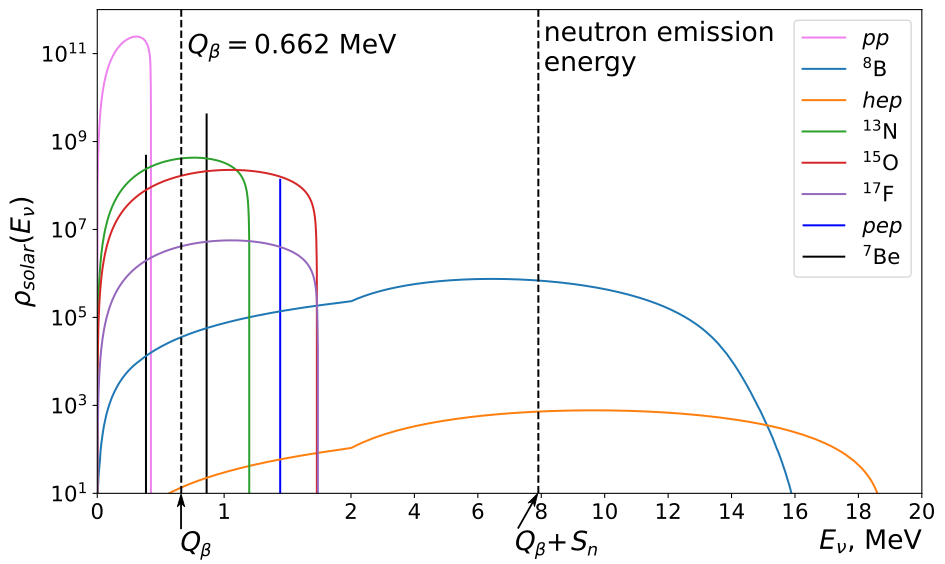}
\caption{The spectrum of solar neutrinos from \cite{Bahcall_2005}. 
The values of $\rho_{solar}(E_{\nu})$ are given in $~cm^{-2}\cdot s^{-1}\cdot \mathrm{MeV}^{-1}$ (for $pep$ and $^7$Be lines in $cm^{-2}\cdot s^{-1}$). 
The threshold energy $Q_{\beta}$ of the reaction
$^{127}$I$(\nu_e,e^-) ^{127}$Xe
and  neutron emission energy $Q_{\beta} + S_n$ of the $^{127}$Xe nucleus are shown. Please note that the scale changes from 2~MeV. $S_n = 7.246$ MeV, $Q_{\beta}= 0.662$ MeV.
}
\label{fig:solar_flux}
\end{figure}

Figure \ref{fig:solar_flux} shows the spectrum of solar neutrinos according to BS05(OP) model \cite{Bahcall_2005}. 
The threshold energy $Q_{\beta} = 662.3 \pm 2.0$~keV of the reaction $^{127}$I$(\nu_e,e^-) ^{127}$Xe and  neutron separation energy $S_n = 7246 \pm 5$~keV \cite{nds.iaea.org} of the nucleus $^{127}$Xe are also shown.
It can be seen that, because of the relatively large value of the energy $Q_{\beta}$ the intense $pp$ neutrinos (reaction $p + p \rightarrow$ $^2\mathrm{H} + \nu_e$) with energies $E(\nu) \le 0.420$~MeV do not contribute to the total capture neutrino spectrum.
Note also that, as can be seen from Figs. \ref{fig:2} and \ref{fig:3}, most of the strength function $S(E)$ of the $^{127}$Xe isotope is located above the energy 1.74~MeV (maximum energy of the CNO neutrinos of the solar cycle), limiting the most intense part of the solar spectrum. 
The $hep$ neutrinos (reaction $^3$He$~+~p \rightarrow ^4$He$~+~e^+ + \nu_e$) 
and boron neutrinos (reaction $^8\mathrm{B} \rightarrow$ $^8\mathrm{Be}^{*} + e^+ +  \nu_e$) could excite the target nuclei to  energies corresponding to the resonant part of the strength function $S(E)$ increasing  the neutrino capture cross-section.

The neutrino-capture rate $R$ (the number of absorbed neutrinos per unit time) relates to the solar neutrino flux and the capture  cross-section by the equation
\begin{equation} 
\label{eq:R_total}
    R = \int_{0}^{E_{max}} \rho_{solar}(E_{\nu}) \sigma_{total}(E_{\nu}) dE_{\nu}
\end{equation}
where, $E_{max} = 16.36$ MeV for boron neutrinos 
and $E_{max} = 18.79$ MeV for $hep$ ones 
\cite{Bahcall_2005}.

The solar-neutrino capture rate is presented in SNU ($1 \times 10^{-36}$ interactions per target atom per second).

\begin{table}[ht]
\caption{  Solar-neutrino  capture  rates R (in  SNU)  for  the  isotope $^{127}$I with fitted experimental strength function (Fig. \ref{fig:2}).
The \textbf{\emph{R-total}} was calculated by using the Fermi-function \cite{Suslov_1968} in two assumptions: with and without neutron emission process in the $^{127}$Xe nucleus. 
For each one the result of exclusion GTR and GTR+PR1 from $S(E)$ is shown. 
}

\centering
\begin{adjustbox}{width=0.9\textwidth}
\small
\begin{tabular}{| c | c | c | c | c | c | c | c |} 
    \hline
\multicolumn{8}{|c|}{\makecell{\textbf{Rate of neutrino capture without neutron emission} \textbf{from the $\mathbf{^{127}}$Хе nucleus}}}  \\
     \hline
      & \textbf{$\mathbf{^{8}}$B} & \textbf{\emph{hep}} & \textbf{$\mathbf{^{13}}$N} & \textbf{$\mathbf{^{15}}$O} & \textbf{$\mathbf{^{17}}$F} &
      \textbf{$\mathbf{^{7}}$Be} &\textbf{Total}  \\
     \hline
    \textbf{\emph{R-total}} & 28.426 & 0.170 & 0.148 & 0.458 & 0.012 & 2.656 & 32.532 \\ \hline
    R without GTR & 8.339 & 0.039 & 0.146 & 0.430 & 0.011 & 2.640 & 12.218 \\ \hline
    R without GTR and PR1 & 5.060 & 0.016 & 0.145 & 0.417 & 0.011 & 2.631 & 8.870 \\
    \hline
\multicolumn{8}{|c|}{\makecell{\textbf{Rate of neutrino capture with neutron emission} \textbf{from the $\mathbf{^{127}}$Хе nucleus}}}  \\
    \hline

    \textbf{\emph{R-total}} & 23.730 & 0.096 & 0.148 & 0.458 & 0.012 & 2.656 & 27.762 \\ \hline
    R without GTR & 8.339 & 0.039 & 0.146 & 0.430 & 0.011 & 2.640 & 12.218 \\ \hline
    R without GTR and PR1 & 5.054 & 0.016 & 0.145 & 0.417 & 0.011 & 2.631 & 8.864 \\
    \hline
\end{tabular}
\end{adjustbox}

\label{table:reaction_rate}
\end{table}

The numerical values of solar-neutrino capture rates $R$ calculated for the reaction $^{127}$I$(\nu_e,e^-) ^{127}$Xe are given in Table \ref{table:reaction_rate} (in SNU).
The calculations were made in two assumptions with and without neutron emission process. Excited states with energies higher than $S_n = 7246 \pm 5$~keV \cite{nds.iaea.org} decay  via neutron emission, whereby the isotope $^{126}$Xe is produced. Difference between this two cases is about $15\%$ to the neutrino capture rate (in \textbf{Total}).

For $^8$B $\sigma_{total} = 4.17/4.99 \times 10^{-42} cm^{-2}$ (with/without neutron emission) is close to the experimental result $(4.3 \pm 0.6)  \times 10^{-42} cm^{-2}$ from \cite{Palarczyk_PhysRevC.59.500}. Our value of $\sigma_{Be} $ is underestimated because of low accuracy of the experimental data at low energies and methodological difficulties in determination of discrete levels width.

The data in Table \ref{table:reaction_rate} (see also Figs. \ref{fig:cross-sec_1} and \ref{fig:cross-sec_2}) shows that the disregard of even GTR leads to a strong reduction of the cross-section and capture rate (approximately by $70\%$) in both calculation modes (with/without neutron emission). Fitting resonances with a Gaussian distribution somewhat redistributes the ratio and strength functions $S_i(E)$ of the contributions of the GTR and pygmy resonances to the cross section, without changing the integral effect from all excitations.
 At solar-neutrino  energies upper than $S_n$,  the number of $^{126}$Xe atoms arises.
The relative amount could be used as an indicator of  the  presence  of hard boron neutrinos  in the  solar spectrum.
From Table \ref{table:reaction_rate}, one can see the amount of the isotope $^{126}$Xe with respect  to $^{127}$Xe  should  be  about  $\approx15\%$.
This  is  of interest for future experiments with a iodine detector, the more so as $^{126}$Xe is a stable isotope released in the form of a gas.

Unfortunately, the accuracy of the experimental strength function measured in known $(p,n)$ reaction  is insufficient to correctly discuss the contribution from beryllium neutrinos. More detailed investigation in  $(^{3} \mathrm{He},t)  $ reaction  with high resolution could allow to clearly estimate the  influence of the discrete components of solar neutrino spectrum.

\section{Conclusions}

We have studied the effect of high-lying resonances  in  the  nuclear   strength  function $S(E)$ for solar-neutrino capture cross-section by $^{127}$Xe  nuclei,  performing  an  analysis  of  available experimental data from the reaction $^{127}$I$(p,n) ^{127}$Xe \cite{Palarczyk_PhysRevC.59.500}.
The new values of the charge-exchange resonances energies are slightly different from those obtained earlier.

On the basis of the self-consistent theory of finite Fermi   systems,   we   have   calculated   the   strength function $S(E)$,   including   the Gamow-Teller  and  analog  resonances,  as  well  as pygmy resonances, which lie lower.
The calculations were  performed  with  the  parameters  of  quasiparticle  local  nucleon–nucleon  interaction  that  were corrected  recently  \cite{Lutostansky2020}.

The  results  of  the calculations  for  the  structure  of  charge-exchange excitations  of  the $^{127}$Xe  nucleus  are  presented,  and the contributions of single-particle $(n-p)$ transitions to the resonance states in question have been found.
The giant Gamow-Teller resonance has been shown to  be  the  most  collective  state. 
The  comparison  of the  calculated  strength  function $S(E)$ with  experimental data has demonstrated good agreement both in  energies  and  in  resonance peak  heights. 
The sum  of  the  calculated  squares  of  matrix  elements for excited states corresponds to the theoretical sum rule with an effective charge of $e_q=0.9$ or $q=0.81$.
This  complies  with  the  observed  \emph{quenching}-effect (underestimation in the sum rule) parameter.

We  have  calculated  the  capture  cross-sections $\sigma(E)$ for  solar  neutrinos  and  have  shown  a  strong effect of the resonance structure on the $\sigma(E)$, especially in the region of high energies.  
Also, we have  analyzed  the  effect  of  each  resonance  on the energy dependence $\sigma(E)$.
We have found that, in calculating $\sigma(E)$, it is necessary to take into account all charge-exchange resonances in the strength function $S(E)$.
The disregard of even the high-lying Gamow–Teller resonance leads to  a  substantial  decrease  of  up to  about $70\%$ in the solar-neutrino  capture  rate  for $^{127}$I  (with  inclusion the  neutron  separation  energy for the $^{127}$Xe  nucleus) owing primarily to high-energy boron neutrinos. 
The contribution from AR resonance to $\sigma(E)$ is negligible ( less than 1\%) for the  reaction $^{127}$I$(\nu_e,e^-) ^{127}$Xe.

We have shown that the inclusion of neutron emission in the consideration reduces the capture rate $R$, especially for boron and $hep$ neutrinos.
For neutrino with energies  higher than $S_n$  the  isotope $^{126}$Xe  arises in the detector medium area. The relative amount of $^{126}$Xe provides an indicator of the high-energy boron and $hep$ neutrinos presence in  the  solar  spectrum.
We have shown that the number of atoms $^{126}$Xe with respect  to $^{127}$Xe  is not small and should be about $17\%$.

This is of interest  for future experiments with iodine detectors, the more so as $^{126}$Xe is a stable isotope released, in this case, in the gas form.

We have investigated the influence of different Fermi-functions to the neutrino capture cross-section $\sigma(E)$. 
It was shown that different approaches gives variation on $\sigma(E)$ to 15\%. 
Nuclear radius gives additional source of uncertainties in determination of Fermi-function. 
Latest investigations of copper \cite{Borzov2020} and potassium \cite{Koszorus:2020mgn} isotopes gives uncertainties in $\delta R \le 4\%$ which provides the same variance in Fermi-function.


Unfortunately, current experimental data for  $^{127}$I were measured only in the $(p, n)$ reaction. To obtain materially more information, it would be proper to have experimental data on the $(^{3} \mathrm{He},t) $ reaction. This allows us to return to the idea of large iodine detectors again and on a new level. An experimental study of this issue will help to determine the neutrino capture cross-section significantly more accurately.

In  our  calculations,  we  disregarded  the  effect  of neutrino  oscillations. 
Neutrino  oscillations  reduce the number of electron neutrinos reaching the Earth because  of  their  transformation  into  other  neutrino flavors. 
However,  it is legitimate to ignore  neutrino oscillations  in  our  case,  where  we  analyze  relative quantities that demonstrate the effect of resonances.

Thus, the article proposes a new type of detector sensitive to the high-energy part of the solar neutrino spectrum.  The convenience of this option could be considered the use of a stable and widespread  in nature $^{127}$I isotope. Also, the chemical inertness of xenon allows to applicate various methods for detecting $^{126}$Xe isotopes.

\section{Acknowledgments}
We are grateful to  D.N. Abdurashitov, F.T. Avignone, A.L. Barabanov, I.N. Borzov, J. Engel,  V.N. Gavrin, L.V. Inzhechik, V.V.Khrushchev, A.Yu. Lutostansky, W. Nazarewicz,  N.B. Shul’gina, M.D. Skorokhvatov, S.V. Tolokonnikov and A.K. Vyborov for stimulating discussions and assistance in the work.

\section{Funding}
This work was supported in part by the Russian Foundation for Basic Research (project no. 18-02-00670); the Division of Neutrino Processes, National Research Center Kurchatov Institute and by the Ministry of Science and Higher Education of the Russian Federation (5-in-100 Program for the Moscow Institute of Physics and Technology (National Research University)).

\printbibliography

\end{document}